\documentclass[aip,
reprint]{revtex4-1}

\usepackage{graphicx}
\usepackage{newtxtext}
\usepackage{newtxmath}
\usepackage{natbib}
\usepackage{hyperref}
\hypersetup{
    colorlinks = true,
    urlcolor   = blue,
    citecolor  = black,
}

\newcommand{\RomanNumeralCaps}[1]
\linenumbers

\usepackage{bm}
\usepackage{subcaption}

\usepackage[detect-none]{siunitx}
\usepackage{placeins}
\usepackage{pgfplots}
\pgfplotsset{compat=1.15}
\newlength\figH
\newlength\figW
\begin{document}


\title{Drop impact on a heated pre-wetted wall: deposition-on-crater splash regime}



\author{Lukas Weimar, Bastian Stumpf,
Jeanette Hussong,
Ilia V. Roisman 
 }
\affiliation{Technische Universität Darmstadt, Institute for Fluid Mechanics and Aerodynamics\\
Peter-Grünberg-Straße 10, 64287 Darmstadt, Germany}

\date{\today}

\begin{abstract}
The impact of a liquid drop with high Reynolds and Weber numbers on a wet solid surface typically results in the emergence, rising, and expansion of a corona-like thin jet. This phenomenon is explained by the propagation of a kinematic discontinuity within the wall film. Conventional theories suggest that the corona-forming liquid jet comprises material from the impacting drop and wall film.

In this study,  the impact of a drop on a wall film is observed using a high-speed video system. Simultaneously, the distribution of the contact temperature at the substrate surface is measured with a high-speed infrared system. The results reveal that heat transfer predominantly occurs within the thin thermal boundary layers in the drop and substrate. Moreover, our experiments show that under our specific conditions, the drop deposits at the base of the crater while only the wall film produces the corona and splashes. Correspondingly, the secondary drops consist only of the heated material of the wall film. This regime has not been previously reported in the literature. 

The validated models for the diameter of the cold spot, the characteristic time, and the contact temperature developed in this study can be potentially useful for reliable modeling of spray cooling. 
\end{abstract}


\maketitle 

\section{Introduction}
\label{sec:intro}

The impact of a single drop on a wetted surface is central to a wide range of applications, including spray painting, spray cooling, inkjet printing, agricultural spraying, soil erosion studies, ice accretion modeling from supercooled droplets, food processing, and more. Several comprehensive reviews have addressed the experimental studies and modeling approaches related to drop impact on thin wall films.\cite{Yarin.2006,Thoroddsen.2008,Liang.2016,Yarin.2017}

The behavior of a drop upon impact is typically characterized by the Reynolds number, $\mathrm{Re} = U D / \nu$, and the Weber number, $\mathrm{We} = \rho U^2 D / \sigma$, where $U$ is the impact velocity, $D$ the drop diameter, and $\nu$, $\rho$, and $\sigma$ represent the fluid's kinematic viscosity, density, and surface tension, respectively. The influence of the wall film is captured by the dimensionless film thickness, $\delta = H_\mathrm{film}/D$.

\begin{figure}
    \centering
    \includegraphics[width=\linewidth]{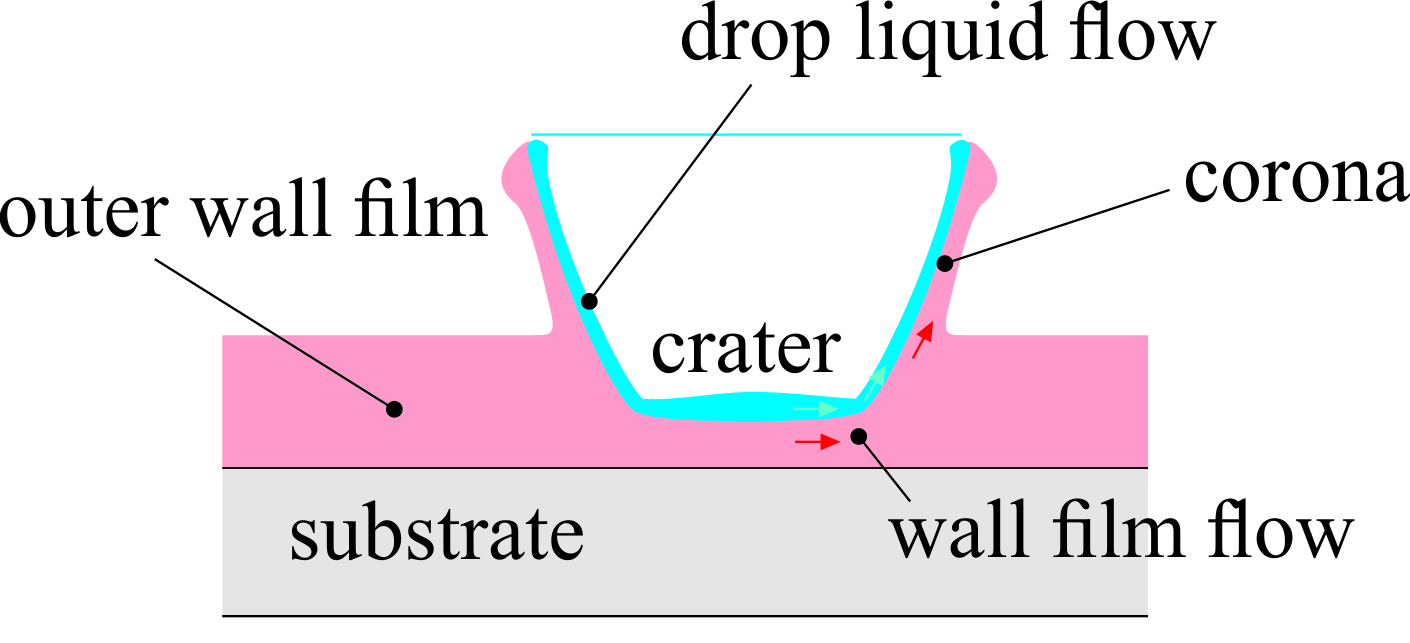}
    \caption{Sketch of the corona formation due to the propagation of the kinematic discontinuity.\cite{Yarin.1995} }
    \label{fig:kinemat}
\end{figure}

At high Reynolds and Weber numbers, drop impact onto a wet substrate is characterized by the initial crater formation in the wall film and by the interaction of the spreading lamella with the outer unperturbed film. This interaction, which is schematically shown in Fig.~\ref{fig:kinemat}, is described as the propagation of a kinematic discontinuity by Yarin and Weiss,\cite{Yarin.1995} leading to the expression for the corona diameter in the form
\begin{equation}\label{eq:dspread}
    d_\mathrm{base} = \beta \sqrt{D U (t+\tau)},
\end{equation}
where the constant $\beta$ depends on the dimensionless wall film thickness $\delta$. Further, more recent theoretical developments for the dynamics of the corona expansion are based on the energy considerations\cite{Gao.2015}  or the analysis of the effects of the viscous boundary layer.\cite{Lamanna.2022}

\begin{figure}
\begin{minipage}{\linewidth}
        \centering
    \includegraphics[width=\linewidth]{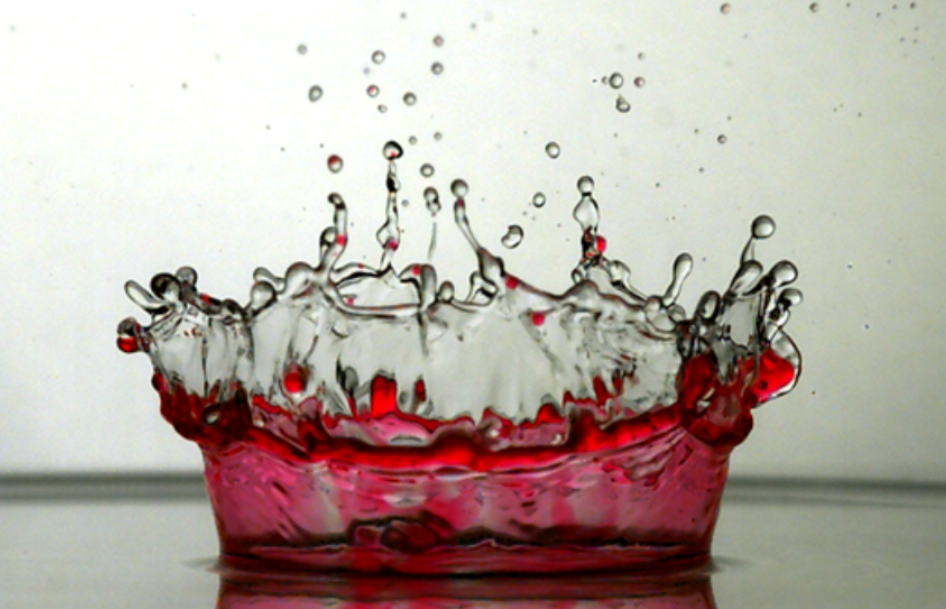}
    \subcaption{Double corona of a dyed water droplet impacting onto a silicone oil S1 wall film:\cite{Kittel.2019}  $D=1.8$ mm, $U=3$ m/s, $H_\mathrm{film}/D=0.2$. }
    \label{fig:Kittel}
\end{minipage}
\begin{minipage}{\linewidth}
        \centering
    \includegraphics[width=\linewidth]{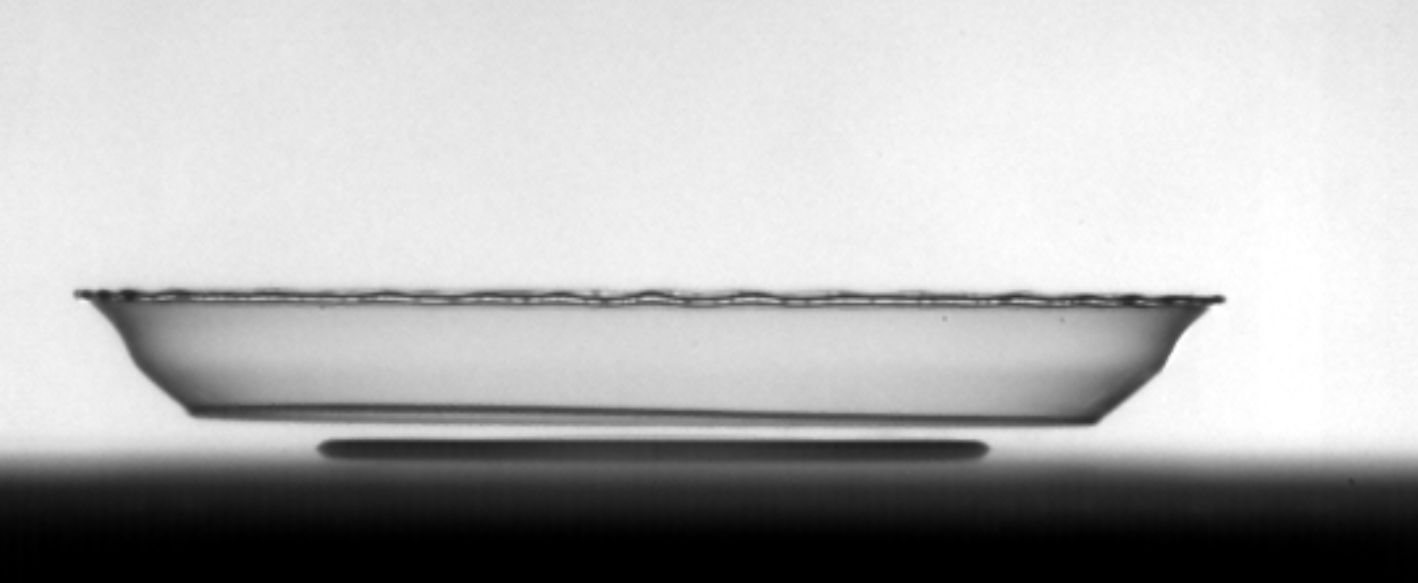}
    \subcaption{Corona detachment after a silicone oil S10 drop impact:\cite{Stumpf.2022} $D=2$ mm, $U=3.2$ m/s, $H_\mathrm{film}/D=0.0185$.}
    \label{fig:Stumpf}
\end{minipage}
\caption{Regimes of drop corona splash.}\label{fig:splashexamples}
\end{figure}

Corona splash occurs when the impact velocity exceeds the splashing threshold.\cite{Cossali.1997,Mundo.1995} This threshold is typically characterized by the dimensionless number $K\equiv \mathrm{We}^{1/2} \mathrm{Re}^{1/4}$. The threshold value of  $K$ depends on the dimensionless film thickness $\delta$, as well as on the viscosity ratio between the drop and film liquids when they differ.\cite{Geppert.2017,Kittel.2018}

Recent observations have revealed intriguing phenomena associated with corona splash. One such phenomenon is the formation of a double corona,\cite{Kittel.2019} which occurs when the height of the liquid sheet formed by the drop is lower than that of the wall-film layer, as illustrated in Fig.~\ref{fig:Kittel}. This effect has been visualized by introducing a dye into the drop liquid. Another notable occurrence, often observed when a drop impacts a wall film significantly thinner than the drop diameter, is the detachment of the corona\cite{Stumpf.2022,stumpf2023drop} from the substrate (see Fig.~\ref{fig:Stumpf}).

Drop impact onto a hot substrate of a temperature significantly exceeding the boiling point is accompanied by various phase change phenomena, leading to the film, nucleate, or transition boiling regime.\cite{Tran.2012,Herbert.2013,Liang.2017,Staszel.2018, Liu.2022,Breitenbach.2018,Schmidt.2021} These phenomena determine the heat flux produced by spray cooling of hot substrate. 

At present, no reliable model for spray cooling has been developed that accurately captures the relevant microphysical phenomena at wall temperatures below the boiling point. It is expected that such a model would require, among other parameters, the characteristic diameter of the impact-induced crater, the characteristic time of crater spreading, and the typical heat flux during this period.  

This experimental study focuses on the hydrodynamics and heat transfer resulting from a drop impacting on a heated, wetted substrate. We found that the size of the cold spot formed by the impact is significantly smaller than the diameter of the resulting corona. This phenomenon is explained by examining the dynamics of the drop spreading at the base of the crater. 

\section{Experimental Methods}

\subsection{Experimental setup}\label{app:exp}

\begin{figure}
    \centering
    \includegraphics[width = \linewidth]{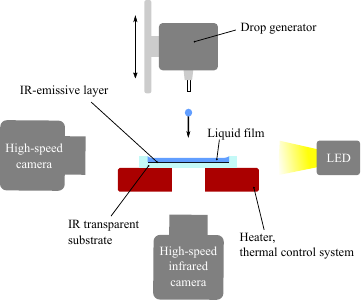} 
    \caption{Schematic illustration of the experimental setup.}
    \label{fig:exp_setup}
\end{figure}

Figure \ref{fig:exp_setup} shows a sketch of the experimental setup used for the drop impact experiments. It comprises a drop generator, a temperature-regulated target substrate with a film thickness sensor, and two high-speed video systems. The horizontal high-speed video system is aimed at capturing the side view of the drop impact, which can be used for observations of the crown produced by the drop impact. The vertical high-speed infrared camera is aimed at providing a bottom view through the IR-transparent sapphire target in the mid-wave infrared (MWIR) spectrum. The images captured by the infrared camera, after careful in-situ system calibration, are used to evaluate the temperature distribution at the substrate interface.

The drop generator consists of a vertical cannula for drop formation. The liquid is supplied by a piezo-electric micropump (\textit{Bartels mp6-liq}). The drop is accelerated toward the target solely under the influence of gravity. The drop diameter and the impact velocity can be varied by using different cannula diameters and adjusting the height of the drop generator, respectively.

A cylindrical sapphire substrate (optically polished, diameter of \SI{50}{mm}, thickness of \SI{3}{mm}), transparent to infrared light (transmissivity for infrared radiation is larger than $\qty{85}{\percent}$), serves as the impact target. The interface exposed to drop impact is coated by a graphite paint (approximately \SI{5}{\micro m} in thickness) with a spray pistol before each experiment. The graphite layer is nontransparent and has a high emissivity in the MWIR spectrum. Radiation emitted by this discrete layer is detected by the IR camera. The substrate is placed on a thermal control system in order to adjust the temperature of the substrate and the liquid film. The thermal control system comprises a cylindrical aluminum component with a flexible heating rod inside, that utilizes Joule heating. The heater has a hole in the center which provides optical access for the IR camera.

The liquid film is applied with a syringe into a cylindrical reservoir that is directly milled into the substrate to prevent the film from leaking. The liquid film thickness is measured with a confocal-chromatic point sensor (\textit{ConfocalDT IFC2421} with \textit{IFS2405-1}) at the center of the film shortly before an experiment. The material properties of silicone oils, used in this study, and their temperature dependencies are either taken from the manufacturer's datasheet\cite{Wacker.2001} if available, deduced from literature\cite{Darhuber.2003,Bhatia.1985,Ricci.1986,Roberts.2017} or verified with measurements. The design of the heater causes the central part of the liquid film to be heated up slightly slower during the heating process caused by the hole in the heater. The corresponding radial Marangoni flow in the liquid film leads to a slight curvature of the film. However, in the center of the film the curvature is negligible, and the liquid film can be considered of constant thickness.

The observation system consists of a high-speed video system (\textit{Photron Fastcam SA-X2}), together with high-power LED (\textit{Veritas Constellation 120E}) for the side-view-investigation of the impact process via shadowgraphy.  The second imaging system comprises a high-speed infrared camera (\textit{Telops Fast-IR M3K}), sensitive to radiation in the MWIR spectrum (wavelength: \num{3}-\SI{5}{\micro m}). A meaningful translation from IR radiation to temperature values requires pixel-wise in-situ calibration before the experiments. For that, a calibration cell with a liquid reservoir is placed on the coated sapphire substrate and heated up homogeneously to different temperatures. While the camera records the radiation values the temperature close to the surface is measured with a Type-J-Thermocouple (accuracy $\pm \SI{1}{\celsius}$). Radiation and temperature are then correlated with a polynomial function based on the Stefan-Boltzmann-Law. The Noise-Equivalent-Temperature-Distribution (NETD), as observed in the thermal images, varies slightly depending on the experimental conditions; however, its value remains approximately $\SI{0.05}{\celsius}$ across all experiments.
Table \ref{tab:cameras} summarizes the relevant parameters of the observation system.

\begin{table}
    \begin{center}
    \def~{\hphantom{0}}
    \begin{ruledtabular}
            \caption{Optical configuration of the observation system: the high-speed video camera and high-speed infrared camera.}
        \label{tab:cameras}
        \begin{tabular}{llll}
        Camera:          & High-speed video& High-speed infrared \\
                   \hline
                Frame rate (\unit{fps}) & $12500$ & $2500$ \\
                Sensor size  (\unit{px^2})  & $1024 \times 1024$ & $320 \times 256$ \\
                Exposure time (\unit{\micro s})& $25$ & \numrange{15}{85} \\
                FOV  (\unit{mm^2})& $23 \times 23$ & $23 \times 18$ \\
        \end{tabular}
    \end{ruledtabular}
    \end{center}
\end{table}

Table \ref{tab:parameter} summarizes the experimental parameters used in this study, where silicone oils with varying viscosities serve as both the drop and film liquids. In each experiment, the drop and film are composed of the same oil; however, temperature differences result in distinct material properties.

\begin{table}
    \begin{center}
    \begin{ruledtabular}
            \caption{Ranges of experimental parameters and material properties}
        \label{tab:parameter}
        \begin{tabular}{ll}
        Parameter & Range  \\
        \hline
        $D$ & $\SI{2.04 \pm 0.05}{mm}$\\
        $U$ & \SI{3.19 \pm 0.04}{m/s}\\
        $\nu_\mathrm{drop}$ & \SIrange{20}{50}{mm^2/s}\\
        $\nu_\mathrm{film}$ & \SIrange{4}{35}{mm^2/s}\\
        $H_\mathrm{film}$ & \SIrange{200}{500}{\micro m}\\
        $T_\mathrm{drop}$ & \SI{24}{\celsius}\\
        $T_\mathrm{film}$ & \SIrange{45}{140}{\celsius}\\
        $\mathrm{Re}_\mathrm{drop}$ & \SIrange{130}{350}{} \\
        $\mathrm{Re}_\mathrm{film}$ & \SIrange{200}{1600}{}  \\
        $\mathrm{We}_\mathrm{drop}$ & \SIrange{880}{1030}{}  \\
        $\mathrm{We}_\mathrm{film}$ & \SIrange{930}{1350}{}  \\
        \end{tabular}
    \end{ruledtabular}
    \end{center}
\end{table}

\subsection{High-speed and infrared observations of drop impact}

\begin{figure*}
\begin{minipage}{\linewidth}
        \centering
  \includegraphics[width=1\linewidth]{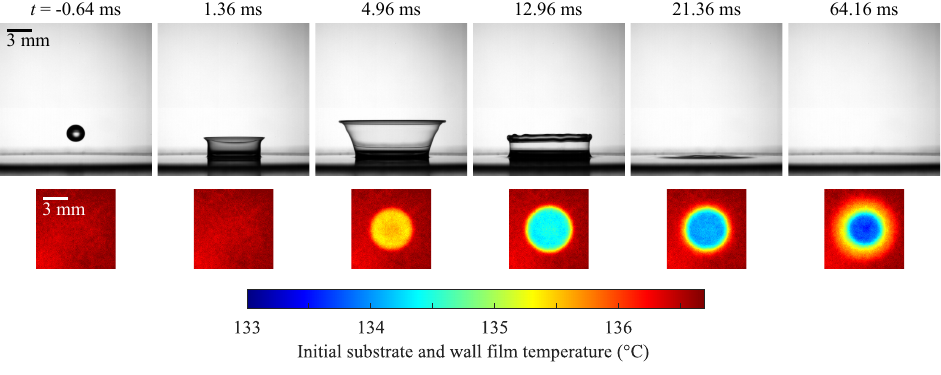} 
    \subcaption{Silicone oil S50,  $\nu_\mathrm{drop}=\SI{50}{mm^2/s}$, $\nu_\mathrm{film}=\SI{10.5}{mm^2/s}$, $H_\mathrm{film}=\SI{343}{\micro m}$, $ T_\mathrm{film}=\SI{136.5}{\celsius}$, $\sigma_\mathrm{drop} \approx \SI{20.8}{mN/m}$, $\sigma_\mathrm{film} \approx \SI{15.4}{mN/m}$.}
    \label{fig:S50_noniso}\vspace{1 cm}
\end{minipage}
\begin{minipage}{\linewidth}
        \centering
    \includegraphics[width=1\linewidth]{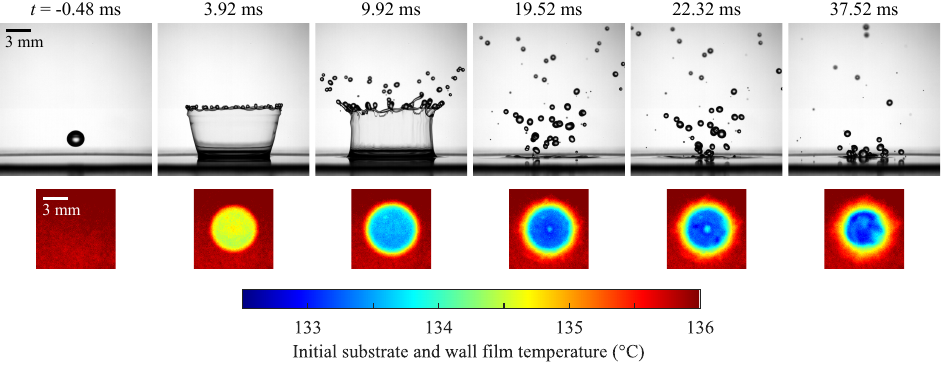} 
    \subcaption{Silicone oil S20, $\nu_\mathrm{drop}=\SI{20}{mm^2/s}$, $\nu_\mathrm{film}=\SI{4.2}{mm^2/s}$, $H_\mathrm{film}=\SI{312}{\micro m}$, $ T_\mathrm{film}=\SI{136}{\celsius}$, $\sigma_\mathrm{drop} \approx \SI{20.5}{mN/m}$, $\sigma_\mathrm{film} \approx \SI{15.1}{mN/m}$.}
    \label{fig:S20_noniso}
\end{minipage}
\caption{Non-isothermal drop impact on a heated film; formation of the corona and a cold spot at the surface. Impact parameters: $D=\SI{2}{mm}$, $U=\SI{3.19}{m/s}$. Impact in (a) leads to the formation of a stable corona, whereas (b) is characterized by splashing behavior. The cold spot remains long after the corona collapses.}\label{fig:noniso}. 
\end{figure*}


In this study, the drop is kept at ambient temperature prior to impact, while the liquid film and substrate are heated. The impact therefore causes local cooling of both the film and the substrate. These effects are observed in Figs.~\ref{fig:S50_noniso} and \ref{fig:S20_noniso}, which show side-view high-speed images and bottom-view IR temperature measurements for a silicone oil drop impacting a heated film of the same oil. The images are labeled with the time $t$ after the instant of impact. In Fig.~\ref{fig:S50_noniso}, the drop impact leads to the formation of an upward-moving, corona-like circular liquid jet that first spreads and then recedes. The example shown in Fig.~\ref{fig:S20_noniso} corresponds to a lower-viscosity drop splash caused by corona instabilities.


The color maps of the infrared images (see the lower rows in Figs.~\ref{fig:S50_noniso} and \ref{fig:S20_noniso})  allow us to identify the appearance and the evolution of a circular cold spot at the substrate interface. Side view and bottom view images are scaled identically to enable direct comparison of the dimensions. Interestingly, in all our experiments, the typical diameter of the cold spot is significantly smaller than that of the corona base. Notably, in the case shown in Figs.~\ref{fig:S50_noniso} ($t = 64.16$ ms), the cold spot persists long after the apparent disappearance of both the corona and the crater in the wall film. A similar persistence is observed in the case depicted in Figs.~\ref{fig:S20_noniso}, where the cold spot remains visible well after corona breakup. Additionally, small, colder spots appear in the infrared images, associated with the deposition of secondary drops onto the blue region representing the cold spot. These findings indicate that the cold spot endures independently of the long-term corona dynamics, whether it breaks up or recedes to cover the crater.


Note that the thickness of the wall film in our experiment is much smaller than the initial drop diameter ($\delta < 0.3$). Therefore, the difference between the diameters of the cold spot and the base of the corona cannot be explained only by the geometrical factors. This surprising result is associated with certain physical phenomena which we try to understand in this study. 

\section{Inertia-dominated drop impact: flow in the spreading drop and wall film }

\subsection{Similarity solution for drop spreading on a dry substrate}

Drop impact onto dry and wetted substrates are, in many respects, fundamentally different phenomena. However, before describing the flow generated by drop impact onto a wetted wall, it is helpful to initially consider the simpler case of impact onto a dry substrate. 

Drop impact generates a radially spreading flow in a thin lamella. The remote asymptotic solution\cite{Yarin.1995}   for the inviscid velocity field  $\{u_r,u_z\}$ in the lamella
\begin{equation}\label{eq:velyarin}
    u_r =r (t+\tau)^{-1},\quad u_z = -2 z(t+\tau)^{-1},
\end{equation}
and the evolution of the lamella thickness\cite{Yarin.1995,Roisman.2009a} 
\begin{equation}\label{eq:hlam}
    h_\mathrm{lamella} \sim D^3 U^{-2} (t+\tau)^{-2},
\end{equation}
are obtained from the mass and momentum balance equations. Here $\tau$ is a constant time shift. This solution is valid for the drop impacts characterized by a high Reynolds number. This solution satisfies exactly the mass and momentum balance equations in a thin spreading lamella. The scaling (\ref{eq:hlam}) has been confirmed by numerous experimental studies, for example, in drop spreading on a convex surface.\cite{bakshi2007investigations,abbot2024spreading} In this case, the Taylor rim formed at the edges of the lamella does not obscure the side view, enabling thickness measurements using the shadowgraphy technique.

Solution (\ref{eq:velyarin}) is valid only when the lamella is much thicker than the viscous boundary layer of the thickness $h_\nu\sim\sqrt{\nu t}$, developed at the instant of impact.  

 A similarity solution of the Navier - Stokes solution for the spreading drop is obtained\cite{Roisman.2009b,Roisman.2010} in the form
\begin{eqnarray}
    u_r &=& g'(\xi)\frac{r}{t}, \quad u_z = - 2 g (\xi) \frac{\sqrt{\nu}}{\sqrt{t}},\\
    g'''&+&2 g g'' + \frac{1}{2} \xi g''+ g' - g'^2=0, \quad \xi = \frac{z}{\sqrt{\nu t}}, \label{eq:simeq}
\end{eqnarray}
  where $g(\xi)$ is the scaled liquid velocity and  $\xi$ is the similarity variable. The solution can be found by numerically integrating the ordinary differential equation (\ref{eq:simeq}), subject to the no-slip and no-penetration conditions at the wall, as well as the requirement that it approaches the outer solution (\ref{eq:velyarin}). The following expression for the lamella thickness is then obtained:\cite{Roisman.2009b}
    \begin{equation}\label{eq:hlamvisc}
    h_\mathrm{lamella} = \frac{\eta D^3}{U^2 t^2} + \frac{4\gamma \sqrt{D U t} }{5 \mathrm{Re}^{1/5}},
    \end{equation}
valid for times $t\gg\tau$.

In the case of drop impact onto a dry, solid, planar surface the computations\cite{Roisman.2009b} of the constants in (\ref{eq:hlamvisc})  yield $\gamma \approx 0.6$, $\eta \approx 0.39$. This is an asymptotic equation for times $t\gg \tau$. 

The shear stress at the wall surface is obtained in the form
\begin{equation}\label{eq:tauw}
    \tau_w = \frac{\partial u_r}{\partial z}_{|z_0} = 1.0354  \frac{ \rho \sqrt{\nu} r}{t^{3/2}}. 
\end{equation}

At long times, when the thickness $h_\nu$ of the viscous boundary layer is of the same order of values as the lamella thickness $h_\mathrm{lamella}$, defined in (\ref{eq:hlamvisc}), the effect of viscosity becomes significant even for high values of Reynolds number. At this stage, the viscous effects lead to the flow decay in the lamella. The lamella thickness starts to deviate significantly from the prediction (\ref{eq:hlamvisc}).  In the case of drop impact onto a dry substrate, the values of the residual film thickness and the characteristic viscous time of drop spreading $t_{\nu,\mathrm{drop}} \sim h_\mathrm{res}^2 \nu^{-1}$ are expressed  in the form\cite{Roisman.2009b} 
\begin{equation}\label{eq:Ddry}
    h_\mathrm{res,\,drop} \approx 0.8 D \mathrm{Re}^{-2/5}, \quad t_{\nu,\mathrm{drop}} =\frac{0.49  D}{U}  \mathrm{Re}^{1/5}. 
\end{equation}

In the case of drop impact onto a dry substrate, the spreading time can be significantly influenced by surface tension $\sigma$ and by the substrate wettability. It has been observed that in the case of drop impact onto a superhydrophobic substrate or in the Leidenfrost regime the collision time,\cite{Senoner2016,renardy2003pyramidal, bird2013reducing, liu2014pancake,richard2002contact,willis2003binary,Schmidt.2021} driven by capillary forces, is expressed as 
\begin{equation}
    t_\sigma \approx \sqrt{\frac{\rho D^3}{\sigma}} = \frac{D}{U}\mathrm{We}^{1/2}.
\end{equation}

The time $t_\sigma$ is determined by the motion of the Taylor rim formed at the lamella edge mainly by surface tension and the force associated with the substrate wettability. This timescale becomes relevant when the duration of rim propagation, including its spreading and receding, is shorter than the development time of the viscous boundary layer. The spreading is determined only by the viscous time $t_{\nu,\mathrm{drop}}$ only if the ratio
\begin{equation}
    \frac{t_{\nu,\mathrm{drop}}}{t_\sigma} = \mathrm{Re}^{1/5} \mathrm{We}^{-1/2},
\end{equation}
is much smaller than unity. A similar conclusion\cite{Clanet.2004} was reached earlier based on different considerations.  For a 2 mm water drop, the condition $\mathrm{Re}^{1/5} \mathrm{We}^{-1/2}\gg 1$ is satisfied when $U\gg 0.7$ m/s. All the experiments in this study satisfy these conditions. Therefore, in the further analysis only the viscosity-based spreading time $t_{\nu,\mathrm{drop}}$ will be considered. 

Moreover, in the case $\mathrm{Re}^{1/5} \mathrm{We}^{-1/2}\gg 1$ the effect of the propagation of the Taylor rim and thus the influence of surface tension on the maximum spreading become negligibly small. The effect of surface tension on drop spreading is also significantly reduced on completely wettable surfaces, where the total capillary force acting on the rim vanishes. This conclusion is based on the numerous experimental data.\cite{Clanet.2004,Roisman.2009b} In this range of impact velocities, the scale for the maximum spreading diameter can be estimated with the help of equation (\ref{eq:Ddry}) and the total volume balance of the impacting drop
\begin{equation}\label{eq:Dspread}
D_\mathrm{spread} \approx 0.9 D \mathrm{Re}^{1/5}. 
\end{equation}

\subsection{Solution for the viscous flow induced by drop impact onto a wetted substrate}\label{sec:viscous}

\begin{figure*}
     \centering
     \begin{subfigure}[t]{0.45\textwidth}
         \centering
         \includegraphics[width=\textwidth]{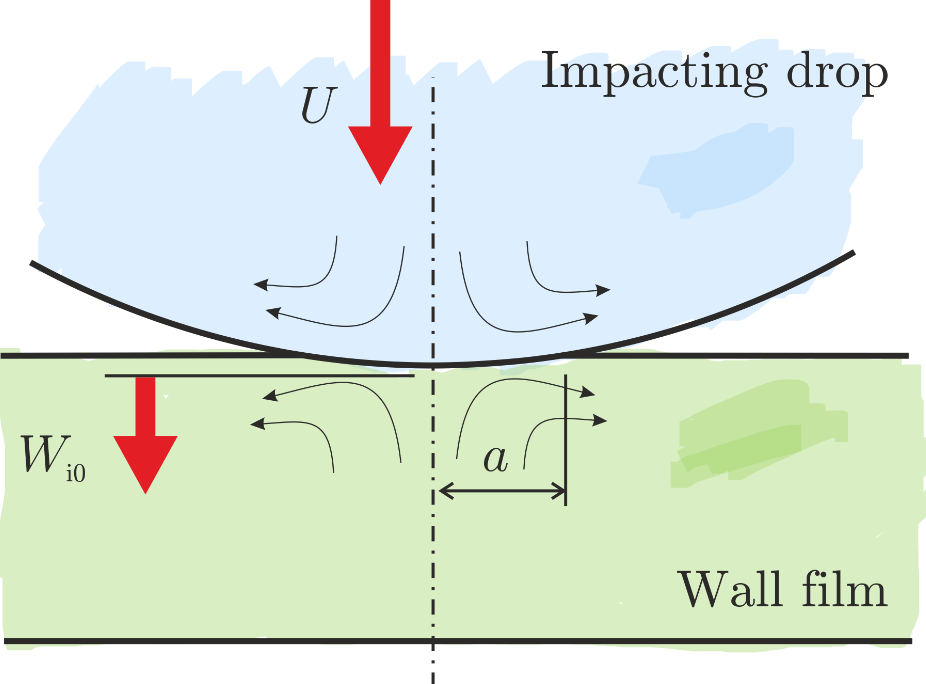}
         \caption{Initial stage of drop penetration}
         \label{fig:stage1}
     \end{subfigure}
     \hfill
     \begin{subfigure}[t]{0.40\textwidth}
         \centering
         \includegraphics[width=\textwidth]{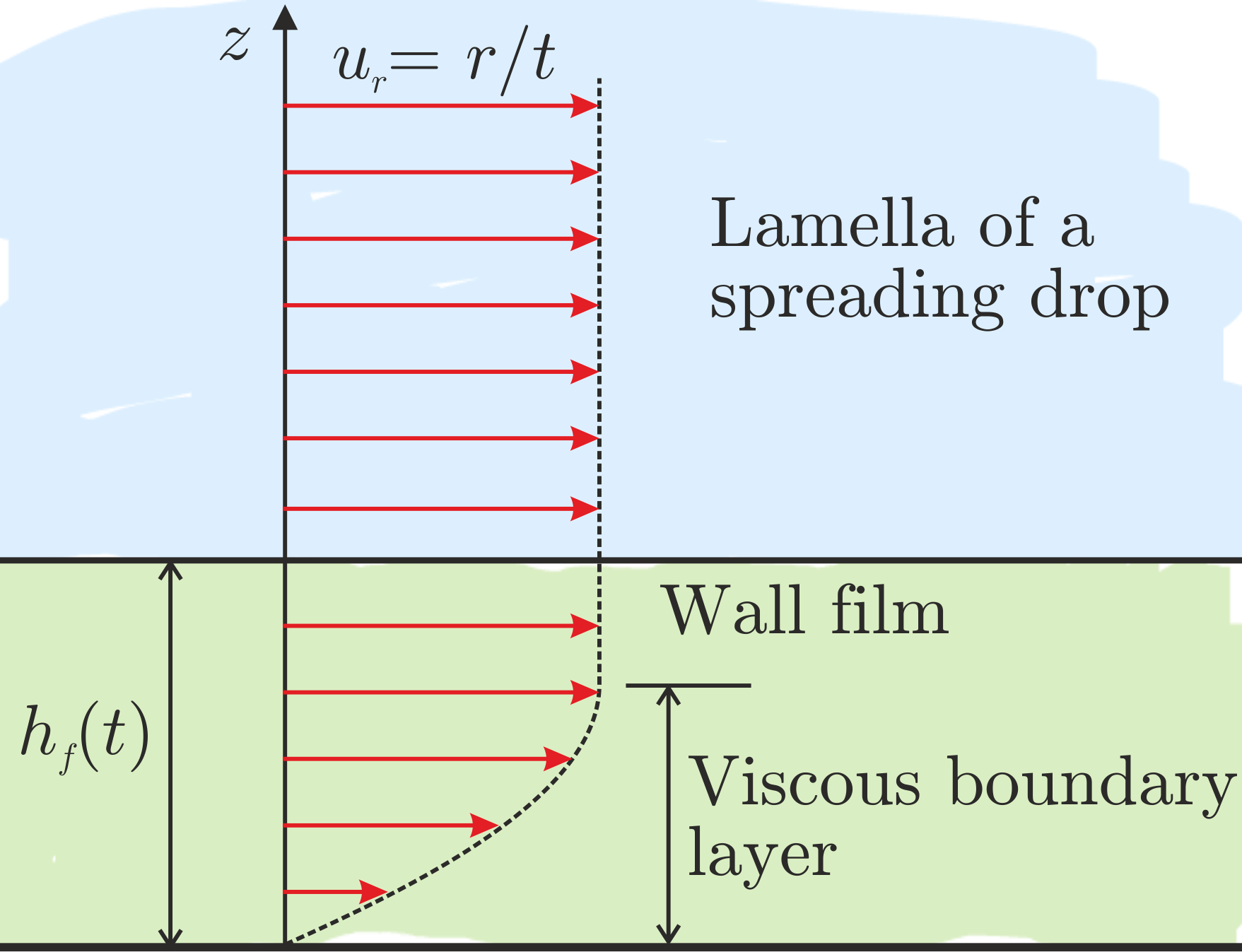}
         \caption{Short times $t<t_{\nu,\mathrm{film}}$}
         \label{fig:stage2}
     \end{subfigure}\\
     \vspace{0.5 cm}
     \begin{subfigure}[t]{0.45\textwidth}
         \centering
         \includegraphics[width=\textwidth]{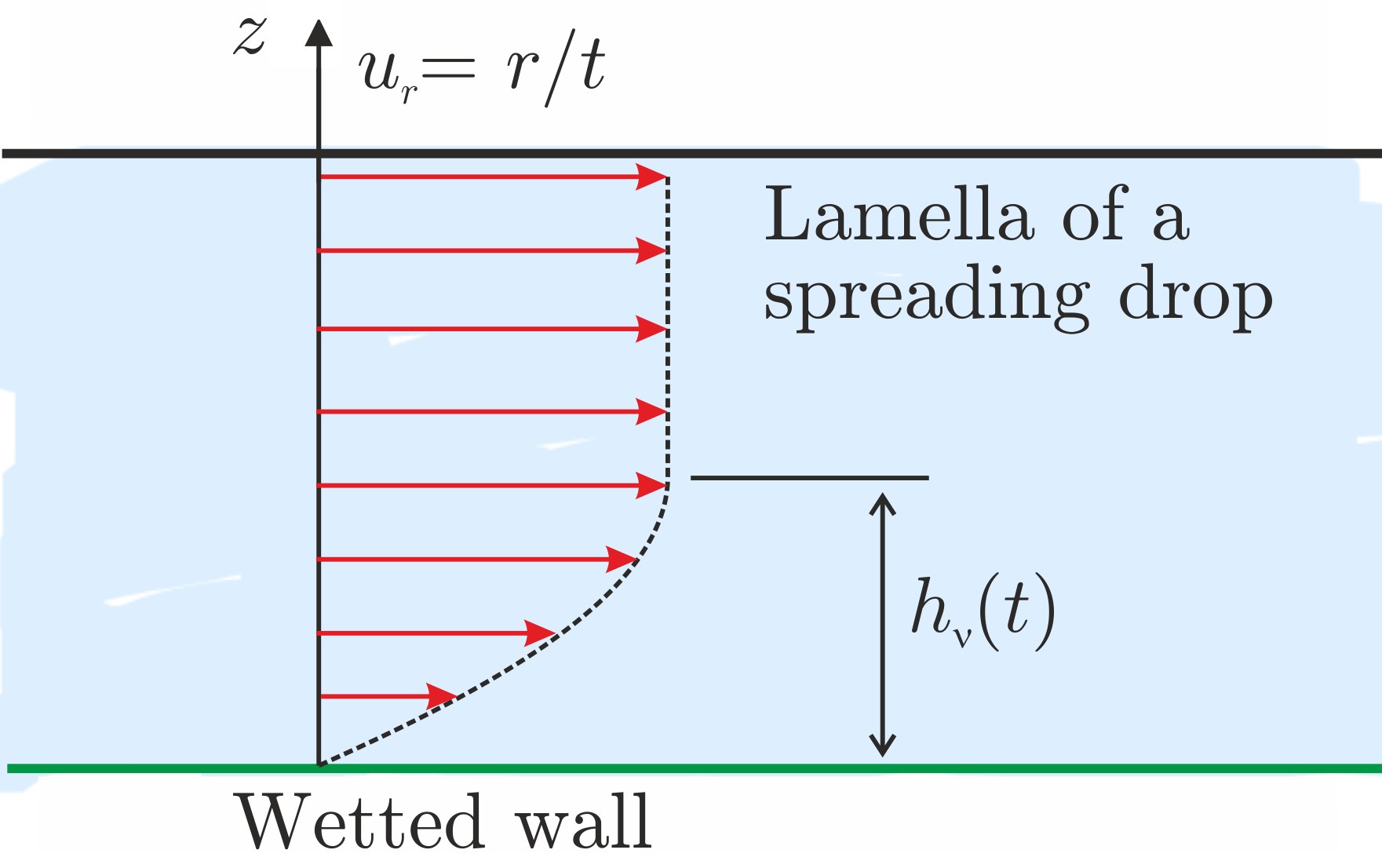}
         \caption{Long times $t\gg t_{\nu,\mathrm{film}}$, outer solution for the lamella of the spreading drop}
         \label{fig:stage3}
     \end{subfigure}
     \hfill     \begin{subfigure}[t]{0.40\textwidth}
         \centering
         \includegraphics[width=\textwidth]{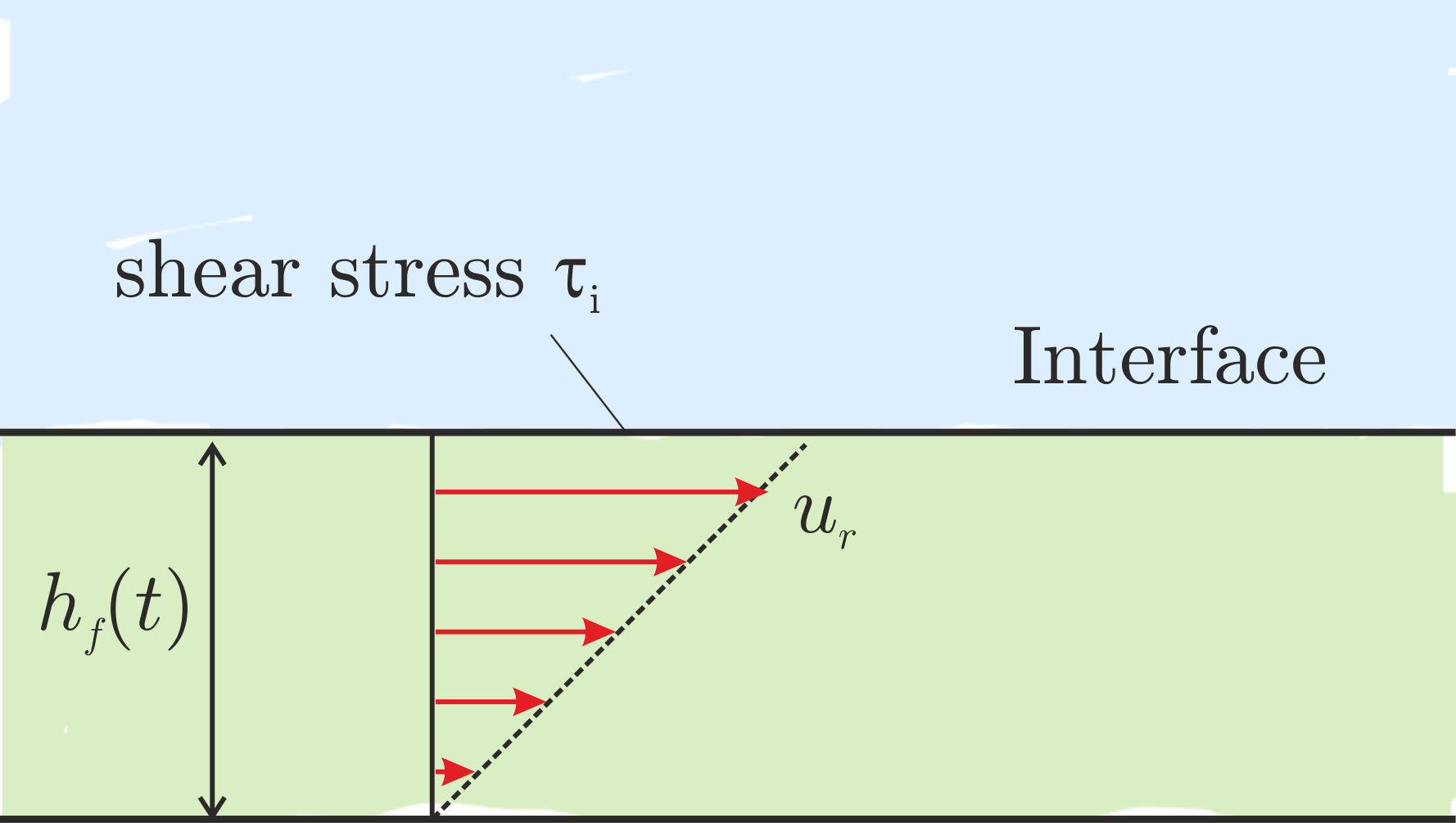}
         \caption{Long times $t\gg t_{\nu,\mathrm{film}}$, inner solution for the flow in the wall film}
         \label{fig:stage4}
     \end{subfigure}\\
          \vspace{0.5 cm}
        \caption{Sketches of the main assumed stages of drop impact, its initial penetration and spreading.\cite{Stumpf.2022} }
        \label{fig:stages}
\end{figure*}

In the case of drop impact onto a wetted substrate the phenomenon can be subdivided into several main stages, shown schematically in Fig~\ref{fig:stages}: (a) initial drop and wall film deformation which may lead to crater formation, (b) expansion of the boundary layer in the wall film at $t<t_{\nu,\mathrm{film}}$, and (c)-(d) the expansion of the viscous boundary layer in the drop region at $t\gg t_{\nu,\mathrm{film}} \ll t  < t_{\nu,\mathrm{drop}}$ characterized by the fully developed viscous flow in the wall film.   At times $t > t_{\nu,\mathrm{drop}}$ the flow in the drop lamella is quickly damped due to viscous effects. The characteristic time  $t_{\nu,\mathrm{film}}$ can be estimated, applying the same considerations to the wall film as in the development of the viscosity-based drop spreading time $t_{\nu,\mathrm{drop}}$, determined in equation (\ref{eq:Ddry})
\begin{eqnarray}\label{eq:stumpftimes}
     t_{\nu,\mathrm{film}} &=& 0.6 D U^{-1} \delta^{2/5}  \mathrm{Re_\mathrm{film}}^{1/5}, \\
     &\mathrm{where} &\quad \delta\equiv \frac{H_\mathrm{film}}{D},\quad \mathrm{Re}_\mathrm{film}\equiv\frac{U D}{\nu_\mathrm{film}}.
\end{eqnarray}

This expression is applicable when the viscosities of the wall film and the impacting drop differ, as in our experiments where the initial temperatures of the drop and the film are unequal. 

The corresponding characteristic viscosity-based wall film thickness $h_{\nu,\mathrm{film}} \sim \sqrt{\nu_\mathrm{film} t_{\nu,\mathrm{film}}}$ at the instant $t_{\nu,\mathrm{film}}$, obtained\cite{Stumpf.2022} using equation (\ref{eq:stumpftimes}), is
\begin{equation}\label{eq:stumpfthick}
    h_{\nu,\mathrm{film}} = 1.49 D \delta^{1/5} \mathrm{Re}_\mathrm{film}^{-2/5}.
\end{equation}

If the initial wall film thickness is much smaller than the drop initial diameter, $\delta \ll 1$, the velocity in the thin wall viscous layer is much smaller than the drop spreading speed. 
 The experiments\cite{Stumpf.2022} show that dependence of the residual film thickness on the initial wall thickness is rather weak if $\delta < 0.1$. Therefore, as a first-order approximation, our solution for the dry substrate (\ref{eq:tauw}) can be used to estimate the shear stress at the interface between the drop and wall film layers at times $t \gg t_{\nu,\mathrm{film}}$. For a given film thickness $h_\mathrm{film}$ and shear stress $\tau_w$, the velocity field in the film can then be estimated from the continuity equation
\begin{eqnarray}
        u_{r\mathrm{film}} &=& \frac{\tau_w z}{\rho_\mathrm{film} \nu_\mathrm{film}} = 1.0354  \frac{ \rho \sqrt{\nu} r z}{\rho_\mathrm{film} \nu_\mathrm{film} t^{3/2}},\\
        u_{z\mathrm{film}} &=& -\frac{1}{r}\int_0^z \frac{\partial(r  u_{r\mathrm{film}})}{\partial r} \mathrm{} z = -\frac{1.0354 \sqrt{\nu } \rho  z^2}{\nu_\mathrm{film} \rho_\mathrm{film}
   t^{3/2}}.\label{eq:uzfilm}
\end{eqnarray}

The evolution of the  thickness $h_\mathrm{film}(t)$ of the wall film is obtained from the condition $h'_\mathrm{film} (t) =  u_{z\mathrm{film}}$ at $z = h_\mathrm{film}(t)$. The corresponding ordinary differential equation is obtained with the help of equation (\ref{eq:uzfilm})
\begin{equation}\label{eq:hevol}
    h_\mathrm{film}'(t) = -  1.0354 \frac{\omega h_\mathrm{film}(t)^2}{\sqrt{\nu_\mathrm{film}}t^{3/2}}, \quad \omega\equiv\frac{\sqrt{\nu}  \rho }{\sqrt{\nu_\mathrm{film}} \rho_\mathrm{film}}.
\end{equation}
which has to be solved subject to the initial conditions $h_\mathrm{film} = h_{\nu,\mathrm{film}}$  at $t=t_{\nu,\mathrm{film}}$. This solution is obtained in the form 
\begin{equation}\label{eq:hfilmont}
    h_\mathrm{film}(t) = \frac{1.49 D {\delta}^{1/5}}{\mathrm{Re}_{\mathrm{film}}^{2/5}
   \left[1+3.86 \omega \left(1-\sqrt{\frac{t_{\nu,\mathrm{film}}}{t}}
   \right)\right]}.
\end{equation}

At large times $t\gg t_{\nu,\mathrm{film}}$ the value $h_\mathrm{film}(t)$ approaches the residual film thickness 
\begin{equation}\label{eq:hfilmres}
    h_\mathrm{res,\,film} = \frac{1.49 D {\delta}^{1/5}}{\mathrm{Re}_{\mathrm{film}}^{2/5}
   \left[1+3.86 \omega\right]}.
\end{equation}

The expression for the total residual thickness of the lamella is obtained  with the help of (\ref{eq:tauw}) and (\ref{eq:hfilmres}) as a sum of the residual drop and wall film layers
\begin{eqnarray}
    h_\mathrm{res} &=& h_\mathrm{res,drop} +h_\mathrm{res,film} = A(\delta) D \mathrm{Re}^{-2/5},\label{eq:hresfinal}\\
    A &=& A_0 + \frac{1.49 \delta^{1/5} \kappa^{2/5}}{1+3.86 \omega},\quad \kappa \equiv\frac{\nu_\mathrm{film}}{\nu}. \label{eq:A}
\end{eqnarray}

\begin{figure}
    \centering
    \includegraphics[width=0.85\linewidth]{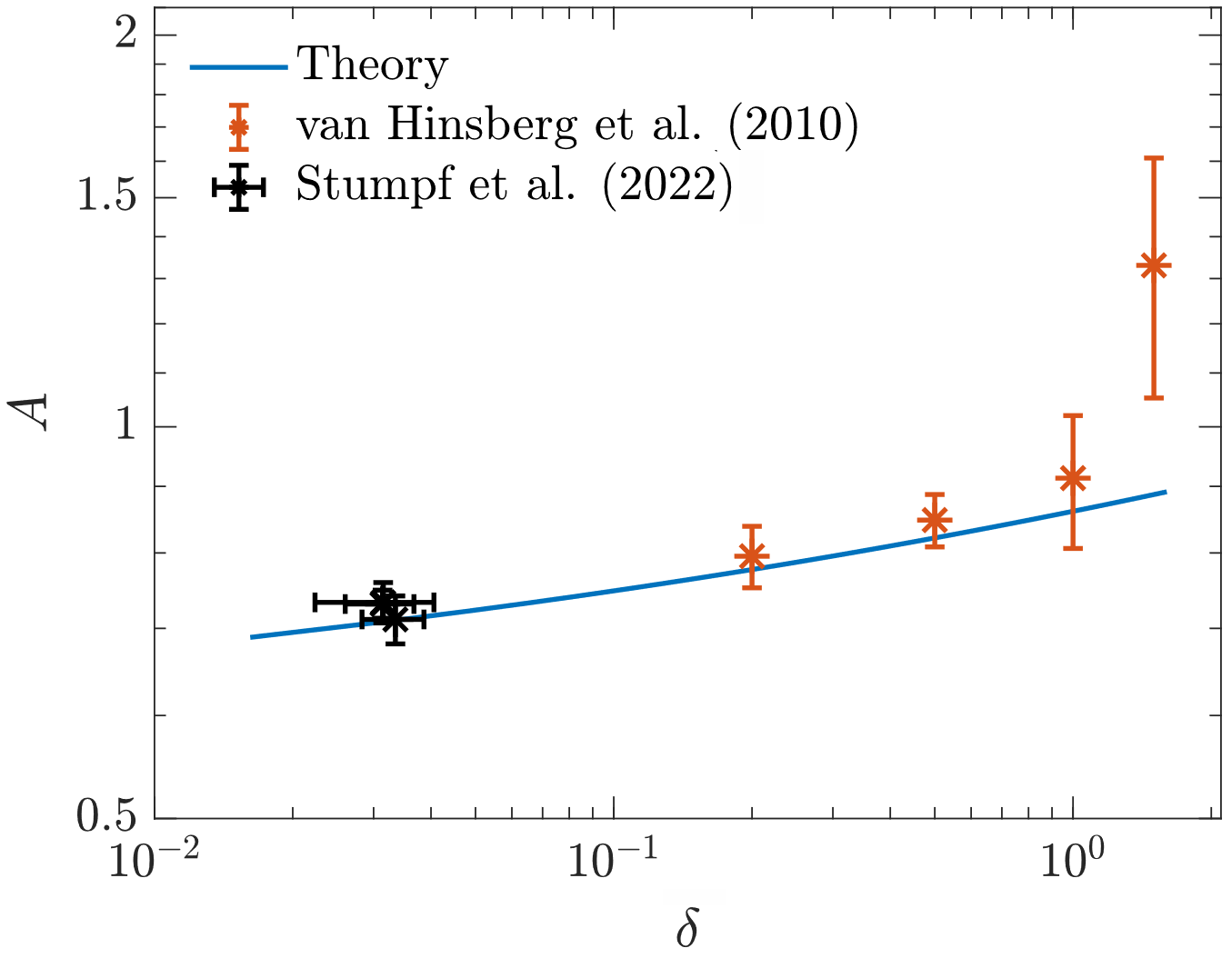}
    \caption{Dependence of the parameter $A$ on the dimensionless wall film thickness $\delta$ for $\kappa =\chi = 1$. The theoretical predictions (\ref{eq:hresfinal}) are shown in comparison with the experimental data from the literatue.\cite{Stumpf.2022,vanHinsberg2010} The Weber number in the experiments varies from 110 to 536, while the Reynolds number ranges from 539 to 8491.}
    \label{fig:Avsdelta}
\end{figure}

The dimensionless constant $A_0 = 0.55$ was obtained\cite{Stumpf.2022} by fitting experimental data for isothermal drop impact onto a liquid film of a different liquid. The theoretical prediction (\ref{eq:hresfinal}) has been validated for liquids of various combinations of drop and wall film viscosities and relatively thin initial wall films with $\delta = 0.02$ and $\delta = 0.06$. The effect of $\delta$ on the values of $A$ is illustrated in Fig.~\ref{fig:Avsdelta}. Although the expression for $h_\mathrm{res}$ is developed strictly for $\delta \ll 1$, the theory (\ref{eq:hresfinal}) still agrees well with experimental data for film thicknesses comparable to the initial drop diameter, up to $\delta = 0.5$. 

\subsection{Thermal effects at early stages of drop impact: time delay}\label{app:a}

\begin{figure}
    \centering
    \includegraphics[width=\linewidth]{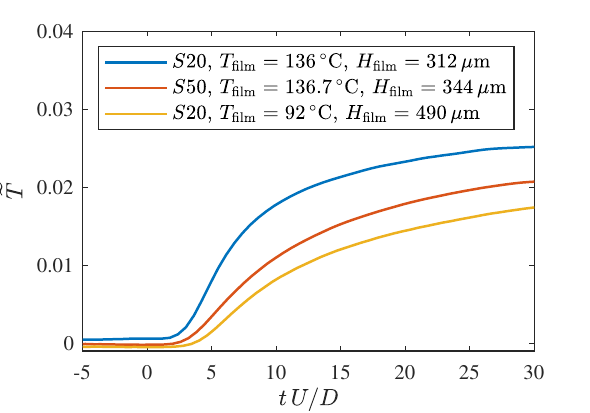}
    \caption{Normalized temperature evolution in center area of the cold spot $\widetilde{T}=(T_\mathrm{center}-T_{\mathrm{substrate}0})/(T_{\mathrm{drop}0}-T_{\mathrm{substrate}0})$ for different experiments with silicone oil under various experimental conditions, showing that the cooling process is delayed.}
    \label{fig:temp_evol_delay}
\end{figure}

The cold spot is always detected after a time delay following the instant of impact. This delay is associated with the propagation time of the thermal boundary layer. It can be detected in Fig.~\ref{fig:temp_evol_delay}, where the evolution of the normalized temperature of the cold spot (averaged over a small central area) is plotted for different exemplary parameters. 

The heat transfer in the drop initially occurs in a thin thermal boundary layer initiated at the drop/film interface. The Prandtl number of the liquids used in the experiments is much higher than unity. Therefore, the thickness of the thermal boundary layer is much smaller than the thickness of the viscous boundary layer. This is why the thermal effects in our experiments can be captured by the HS IR camera only after some notable delay after the drop impact. This delay is determined by the time required for the thermal effects to reach the substrate. 

The time evolution of the thickness of the thermal boundary layer can be estimated with $h_{\Theta}=\sqrt{{\alpha_\mathrm{film}}t}/c$, where $\alpha_\mathrm{film}$ is the thermal diffusivity of the liquid film, $t$ is the time and $c$ is a fitting parameter in the order of unity. 

Therefore, the time delay $t_\mathrm{delay}$, associated with the appearance of the thermal effects at the drop/substrate interface can be estimated from the condition $h_{\Theta}=h_\mathrm{film}$, where $h_\mathrm{film}$ is the lamella thickness, defined in (\ref{eq:hfilmont}). The time delay can be therefore obtained as a root of the equation
\begin{equation}
    \sqrt{{\alpha_\mathrm{film}}t_\mathrm{delay}} =  \frac{1.49 D {\delta}^{1/5}}{\mathrm{Re}_{\mathrm{film}}^{2/5}
   \left[1+3.86\omega\left(1-\sqrt{\frac{t_{\nu,\mathrm{film}}}{t_\mathrm{delay}}}
   \right)\right]}.
\end{equation}
The solution for time delay is obtained using (\ref{eq:stumpftimes}) in the form
\begin{eqnarray}
    t_\mathrm{delay}& =& 0.6 \delta^{2/5}  \mathrm{Re}_\mathrm{film}^{2/5} \frac{D}{U} 
\left( \frac{3.86\omega + 1.93 c \mathrm{Pr}_\mathrm{film}^{1/2} }{3.86\omega+1}\right)^2, \label{eq:tdelay}\\
    \mathrm{Pr}_\mathrm{film} &\equiv& \frac{\nu_\mathrm{film}}{\alpha_\mathrm{film}},
\end{eqnarray}
where $\mathrm{Pr}_\mathrm{film}$ is the Prandtl number of the wall film liquid. 

\begin{figure}
    \centering
    \includegraphics[width=\linewidth]{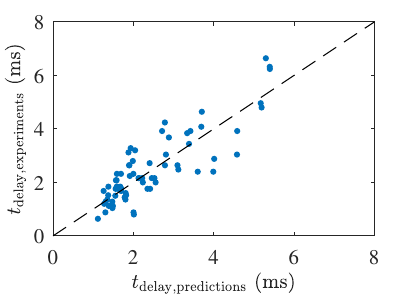} 
    \caption{Measured time delay as a function of the theoretical predictions (\ref{eq:tdelay}) with a fitting parameter of $c = 0.36$.}
    \label{fig:time delay}
\end{figure}

The theoretically predicted time delay is compared with experimental data shown in Fig.~\ref{fig:time delay}, where various drop impact parameters (such as initial wall film thickness, film temperature, and viscosity) are varied. For the fitting parameter $c=0.36$ (standard fitting error $\sigma \approx 0.0086$), good agreement between the experiments and the model, indicated by the straight dashed line, is observed. This result allows for the estimation of the thermal boundary layer thickness in the form $h_\theta \approx 2.8 \sqrt{\alpha_\mathrm{film} t}$, which is physically reasonable.  Some scatter in the data is attributed to the relatively low temporal resolution of the IR camera, which has a frame exposure time of $\SI{0.4}{ms}$. Consequently, significant cooling of the wetted interface begins once the drop lamella is fully heated.

\subsection{Fully developed viscous flow at late times $t\gg t_{\nu,\mathrm{drop}}$}\label{sec:fullydev}

The drop spreading on the wall film may be slightly influenced by the flow, generated in this film. This effect has been neglected in the considerations in \S \ref{sec:viscous}. At large times, corresponding to the viscous flow in the thin drop and wall film layers, the main additional influencing parameter (in comparison to the case of drop impact onto a dry solid wall) is the interfacial velocity $u_\mathrm{int}$. The value of this interfacial velocity can be estimated from the condition of the continuity of the viscous shear stresses at the interface. The shear stress in each layer at larger times can be estimated as $\tau_\mathrm{drop} \sim \mu_\mathrm{drop} (u_\mathrm{drop}-u_\mathrm{int})/h_\mathrm{res}$ and $\tau_\mathrm{film} \sim \mu_\mathrm{film} u_\mathrm{int}/h_\mathrm{res,\,film}$, respectively. Here $u_\mathrm{drop}$ is the characteristic velocity in the drop while $\mu$ is the dynamic viscosity of either drop or film. 

The characteristic thicknesses of the drop and the wall film are determined in the equations (\ref{eq:hresfinal}) and (\ref{eq:A}). 

 The continuity of the shear stresses in the limit $u_\mathrm{int}\ll u_\mathrm{drop}$, valid for small values of $\delta$ yields the following condition
\begin{equation}\label{eq:chi}
u_\mathrm{int}\sim \chi u_\mathrm{drop}, \quad \quad \chi = \frac{\delta^{1/5} \omega}{(1+3.86 \omega)\kappa^{1/10}},
\end{equation}
where the parameters $\omega$ and $\kappa$ are defined in (\ref{eq:hevol}) and (\ref{eq:A}). 
Ratio $\chi$ is a dimensionless parameter that characterizes the influence of the wall film on the spreading of the drop. The properties of the drop and wall film liquids differ from each other due to temperature differences or if the drop and film liquids are different.  

\section{Drop Deposition-on-Crater Phenomenon}

\begin{figure}
    \centering
    \includegraphics[width = 0.8\linewidth]{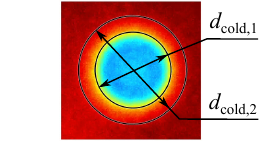}
    \caption{Definition of the characteristic diameters of the cold spot, captured by the IR system: the inner $d_\mathrm{cold,1}$ and the outer $d_\mathrm{cold,2}$. }
    \label{fig:D_def}
\end{figure}

To characterize the geometry of the cold spot caused by drop impact onto an initially heated wall film, two of its characteristic diameters are determined, the inner $d_\mathrm{cold,1}$ and the outer $d_\mathrm{cold,2}$, as shown in Fig.~\ref{fig:D_def}. The inner diameter $d_\mathrm{cold,1}$ is associated with the maximum radial temperature gradient, and the outer diameter $d_\mathrm{cold,2}$ is determined by a low dimensionless temperature threshold (for the shown experiments a value of $0.08$ is used) to obtain the circular area that is influenced by cooling.

%
%


\begin{figure*}
    \centering
    \includegraphics{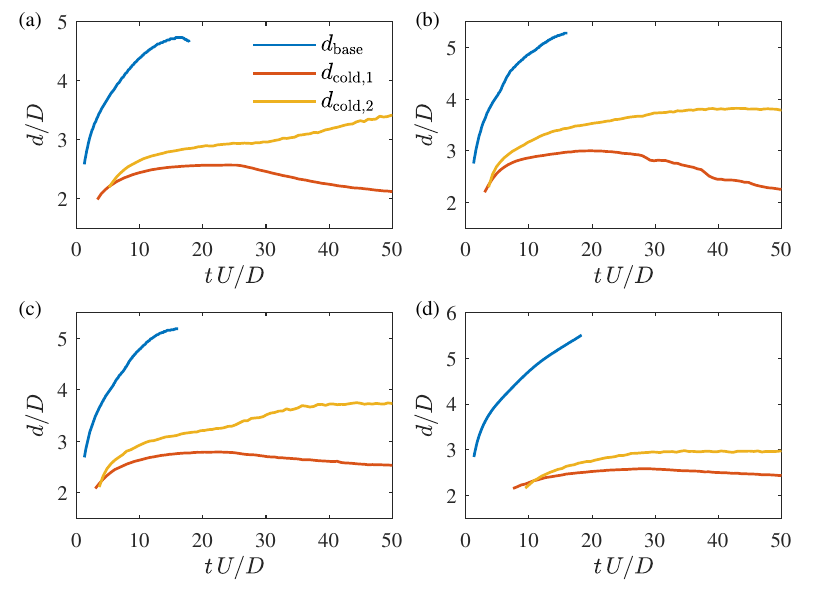}
    \caption{Different exemplary time evolutions of the crown base diameter compared with the characteristic diameters of the cold spot, (a): $S50$, $T_\mathrm{film}=\SI{136.5}{\celsius}$, $H_\mathrm{film}=\SI{343}{\micro m}$ from Fig.~\ref{fig:S50_noniso}, (b): $S20$, $T_\mathrm{film}=\SI{136}{\celsius}$, $H_\mathrm{film}=\SI{312}{\micro m}$ from Fig.~\ref{fig:S20_noniso}, (c): $S20$, $T_\mathrm{film}=\SI{91.5}{\celsius}$, $H_\mathrm{film}=\SI{316}{\micro m}$, (d): $S50$, $T_\mathrm{film}=\SI{92}{\celsius}$, $H_\mathrm{film}=\SI{500}{\micro m}$.}
    \label{fig:D_evol}
\end{figure*}

In Fig.~\ref{fig:D_evol} the two characteristic diameters of the cold region are compared with the diameter of the corona base. The typical size of the cold spot in our experiments is much smaller than the corona base diameter. The inner diameter of the cold spot $d_\mathrm{cold,1}$ initially increases and then, at some time ($t_\mathrm{max} U/D \approx 25$ in the example in  Fig.~\ref{fig:D_evol}a) reaches the local maximum $d_\mathrm{cold,max}$. At larger times the transport processes in the liquid regions lead to the smearing of the boundary of the cold region. The measurements show that $d_\mathrm{cold,2}-d_\mathrm{cold,1} \approx 4 \sqrt{\nu_\mathrm{film} (t- t_\mathrm{max})}$, which indicates that the radial expansion of the viscous boundary layer causes the smearing. 

One possible explanation for the significantly smaller diameter of the cold spot compared to the corona is the specific drop spreading regime in which the drop spreads along the bottom of the crater. Since the residual thickness of the drop lamella is determined by Eq.~(\ref{eq:hresfinal}), the drop spreading diameter can be estimated from the drop mass balance and the empirical coefficient $A_0 = 0.55$, in the form
\begin{equation}
    D_\mathrm{spread} = \sqrt{\frac{2}{3 A_0}} D \mathrm{Re}^{1/5} \approx 1.1  D \mathrm{Re}^{1/5}.
\end{equation}

\begin{figure}
    \centering
    \includegraphics{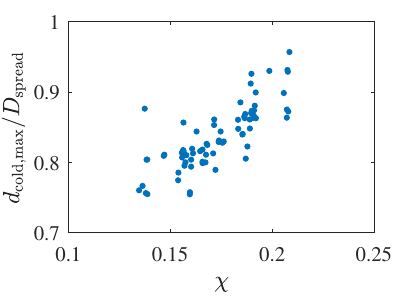}
    \caption{Maximum expansion diameter of the cold spot, scaled by the diameter $D_\mathrm{spread}$ as a function of the dimensionless parameter $\chi$, defined in equation (\ref{eq:chi}), for different experiments in which the parameters film thickness, film temperature and viscosity are varied.}
    \label{fig:D_ir_max}
\end{figure}

In Fig.~\ref{fig:D_ir_max} the measured values of the maximum expansion diameter of the cold spot $d_\mathrm{cold,max}$, scaled by the drop spreading diameter $D_\mathrm{spread}$, defined in equation (\ref{eq:Ddry}) are shown as a function of the dimensionless parameter $\chi$, defined in equation (\ref{eq:chi}). In \S \ref{sec:fullydev} we have assumed that this dimensionless parameter captures the dominant influence of the wall film on the flow within the drop lamella. Not only is the ratio $d_\mathrm{cold,max}/D_\mathrm{spread}$ rather close to unity but it also slightly and monotonically increases with the parameter $\chi$. The apparent spreading diameter is slightly smaller than $D_\mathrm{spread}$ due to the viscous terms at the lamella edge. 

This result confirms our assumption that the cold spot in our experiments is associated with the deposited liquid drop. Its maximum diameter is much smaller than that of the crater, and the drop liquid therefore does not enter the corona. In the case of a splash, the secondary drops consist exclusively of the wall film liquid.

\begin{figure}
    \centering
    \includegraphics[width=\linewidth]{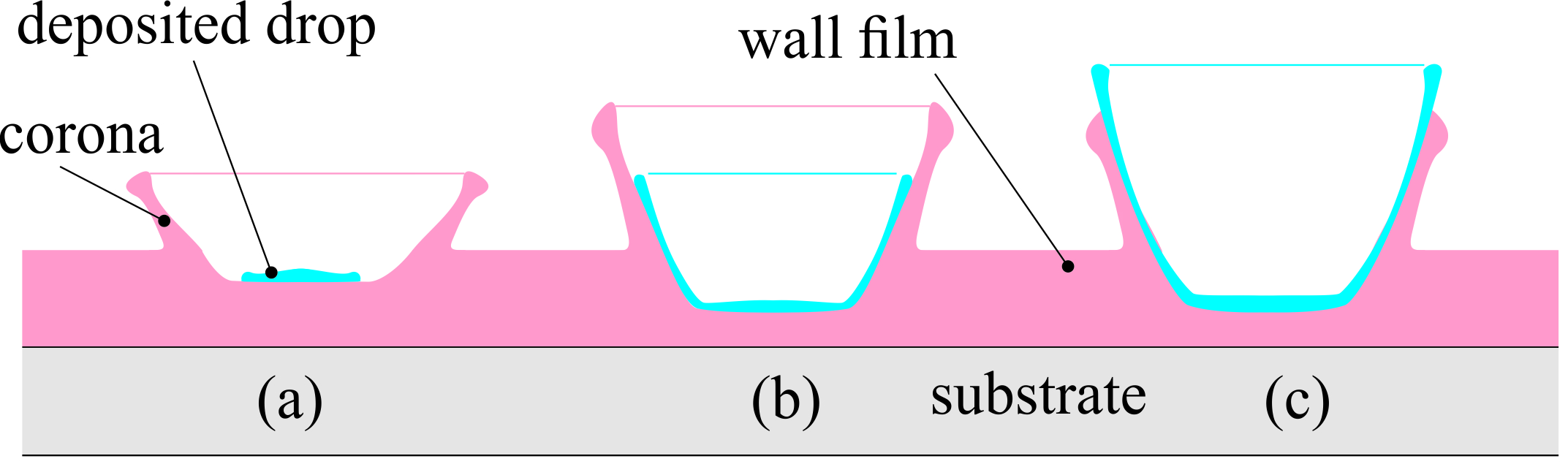}
    \caption{Sketches of various regimes of corona splash: (a) drop deposition-on-crater regime, (b) wall-film-dominant corona, and (c) drop-dominant corona.}
    \label{fig:sketches}
\end{figure}

The \emph{deposition-on-crater regime} investigated in this study is shown schematically in Fig.~\ref{fig:sketches}a. The phenomenon of drop deposition on the crater in the wall film also explains the observations of the residual on the film whose diameter is significantly smaller than the corona diameter, shown in Fig.~\ref{fig:Stumpf}. Moreover, we expect that at some impact conditions the diameter $D_\mathrm{spread}$ can exceed the diameter of the corona. The corresponding \emph{wall-film-dominant corona regime}, observed in the experiments in Fig.~\ref{fig:Kittel}, is shown schematically in Fig.~\ref{fig:sketches}b. It is logical to assume that, under specific impact parameters, the \emph{drop-dominant corona regime}, shown schematically in Fig.~\ref{fig:sketches}c, can also be observed, though not yet.

\begin{figure*}
    \centering
    \includegraphics[width=\linewidth]{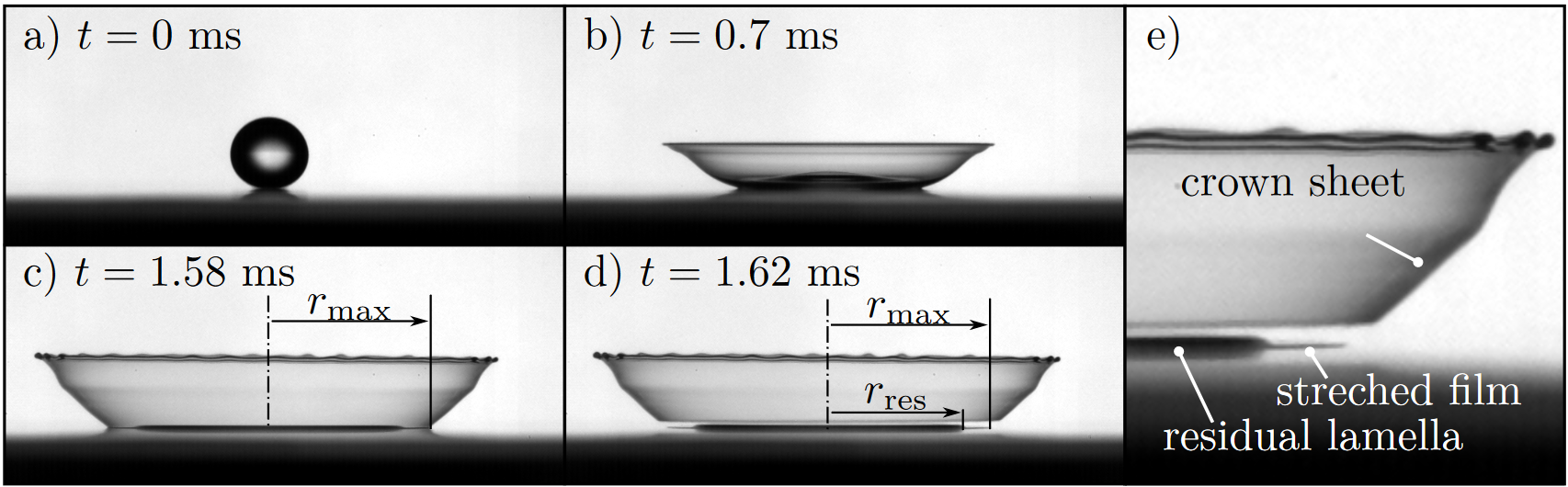}
    \caption{Temporal evolution of corona with the maximum corona radius $r_\mathrm{max}$ defined in c) and the radius of the residual film $r_\mathrm{res}$  defined in d). Impact parameters: Drop and film liquid is S10, $U$ = 3.2 m/s, $D$ = 2 mm, $H_\mathrm{film}$ = 52 \textmu m. A zoomed-in section of (d), presented in (e), shows the geometry of the lamella near the crown base shortly after detachment.}
    \label{fig:coronadepos}
\end{figure*}

Evidence for the existence of the \emph{deposition-on-crater regime} is presented in Fig.~\ref{fig:coronadepos}. In this case, the geometry of the residual layer is visible due to the sudden detachment of the corona. The radius $r_\mathrm{res}$ of the resulting layer is smaller than the maximum radius $r_\mathrm{max}$ of the corona.

The corona splashing regime is a key factor influencing the efficiency of spray cooling. This efficiency is reduced in drop-dominant coronas, where splashing leads to the loss of cold drops that would otherwise contribute to wall cooling. Conversely, in the drop deposition-on-crater regime, the entire mass of cold drops in the spray avoids splashing, instead depositing on the substrate surface and contributing effectively to the heat flux.

A necessary condition for the occurrence of the drop deposition-on-crater regime is that the drop's spreading diameter remains significantly smaller than the corona diameter. The dynamics of the corona are governed by inertial and viscous effects, surface tension, gravity\cite{vanHinsberg2010}, and geometrical factors such as the initial dimensionless film thickness. 

\begin{figure*}
    \centering
    \includegraphics[width=\linewidth]{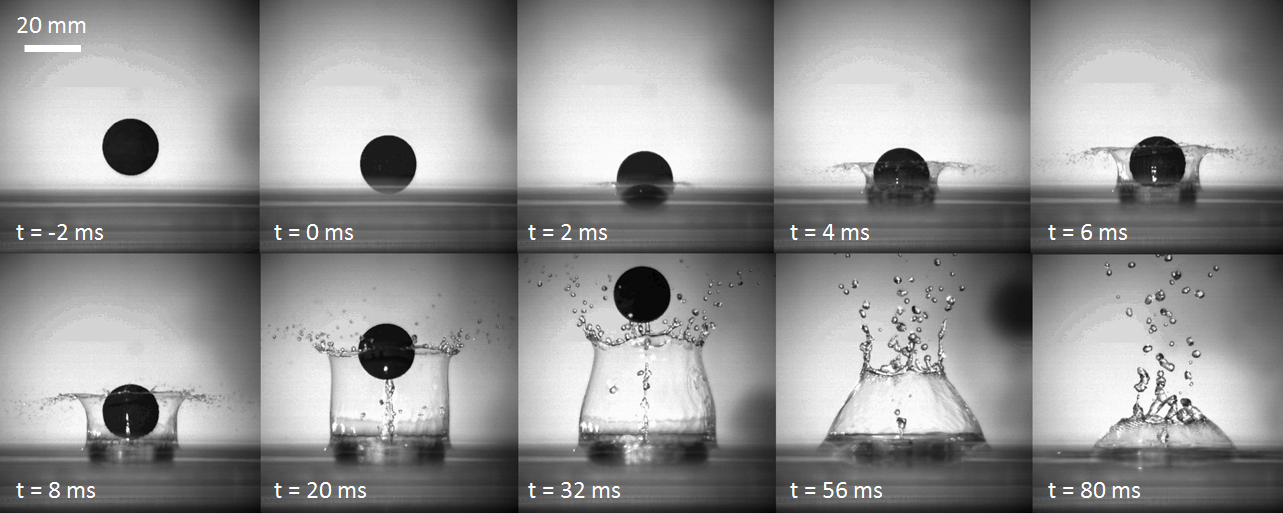}
    \caption{Impact of an EPDM sphere\cite{Hauk_solidparticle} onto a prewetted substrate with $D$ = 20 mm, $U$ = 3.2 m/s and $H_\mathrm{film}$ = 5 mm. }
    \label{fig:solidsphere}
\end{figure*}

It also remains unclear how the corona propagates when the internal flows within the drop and the wall film are damped by viscosity. The most illustrative example of this phenomenon is the wall film splash following the impact of a solid spherical particle, as shown in Fig.~\ref{fig:solidsphere}. The corona continues to spread driven by the inertial terms in the liquid sheet even after the rebound of the EPDM particle. 

Developing a reliable model for the corresponding threshold criterion, based on impact parameters and liquid properties, still requires additional experimental data, particularly on the kinematics of corona expansion. This represents an interesting direction for future research.

\section{Conclusions}

The results of this study demonstrate that the flow and heat transfer during drop impact can be effectively described by the evolution of viscous and thermal boundary layers in both the liquid and solid phases. Analysis of the corona morphology highlights the need to account for distinct interaction regimes between the drop and the wall film. Notably, the deposition-on-crater regime, investigated here, appears to be the most efficient for spray cooling, as it leads to splashing of only the heated wall film while the colder drop is fully deposited. These insights may contribute to the development of more reliable spray cooling models. Additionally, thermal scaling enables the estimation of the contact temperature at the wall interface.

\acknowledgments{The authors gratefully acknowledge the financial support of this work by the Deutsche Forschungsgemeinschaft (DFG) in the framework of TRR 150 (Project-ID: 23726738). I.R and J.H. have been also partially supported by the joint DFG/FWF Collaborative Research Centre CREATOR (DFG: Project-ID 492661287/TRR 361; FWF: 10.55776/F90, subproject B03). The authors thank the anonymous reviewers of \emph{Physical Review Letters} and \emph{Journal of Fluid Mechanics} for their constructive comments, which significantly improved the manuscript.}

\section*{Declaration of Interests} The authors report no conflict of interest.


\section*{References}
\bibliography{bibliography,resid,thermosuperlit}

\end{document}